\documentclass[prl,twocolumn,showpacs,preprintnumbers,amsmath,amssymb,superscriptaddress]{revtex4}

\usepackage{graphicx}
\usepackage{dcolumn}
\usepackage{bm}

\begin{document}

\title{Do we live in a ``Dirac-Milne'' Universe?}

\author{Aur{\'e}lien Benoit-L{\'e}vy}
\email{benoitle@csnsm.in2p3.fr}

\affiliation{CEA/IRFU/SPP, 91191 Gif-sur-Yvette, France}
\affiliation{Universit{\'e} Paris-Sud \& CNRS/IN2P3/CSNSM, 91405 Orsay, France}
\author{Gabriel Chardin}
\email{chardin@csnsm.in2p3.fr}

 \affiliation{Universit{\'e} Paris-Sud \& CNRS/IN2P3/CSNSM, 91405 Orsay, France}
\date{\today}

\begin{abstract}
 The $\Lambda$-CDM standard model, although an excellent parametrization of the present cosmological data, requires two as yet unobserved components, Dark Matter and Dark Energy, for more than 95\% of the Universe, and a high level of fine-tuning. Faced to this unsatisfactory situation, we study an unconventional cosmology, the Dirac-Milne universe, a matter-antimatter symmetric cosmology, in which antimatter is supposed to present a negative active gravitational mass. We show that this universe remarkably satisfies the cosmological tests for the age of the Universe, Big-Bang Nucleosynthesis and Type Ia Supernovae data. Most surprisingly, it also provides the degree scale for the first acoustic peak of the Cosmological Microwave Background. This simple model, without any adjustable parameter or need for Dark Matter or Dark Energy, is a reminder that we should look for simpler and more motivated cosmological models than the present $\Lambda$-CDM standard model.
\end{abstract}

\maketitle

The Concordance Model of the $\Lambda$-CDM universe represents an impressive fit of the present cosmological data, but also represents a theoretical challenge as it requires a high level of fine-tuning. Faced to this uncomfortable situation, we study the unconventional cosmology of a symmetric universe, {\it i.e.} containing equal quantities of matter and antimatter, where antimatter is supposed to present a negative active gravitational mass. At large distances, above the characteristic length of the matter-antimatter emulsion, this universe behaves as if it were gravitationally empty and is therefore neither accelerated nor decelerated. It is therefore characterized by a scale factor linearly dependent in time. At cosmological scales, this cosmology is adequately described by a Milne universe (also sometimes called a ``coasting'' universe), and in the following, we will refer to it as the Dirac-Milne universe, by analogy with the Dirac sea similarly involving negative and positive energy states.  For the purpose of this article, we will mostly use the property of linear dependance in time of the scale factor.
After a brief summary of our motivations,  we will show how this universe provides a scenario for the production of light nuclei during the Big-Bang Nucleosynthesis and how the Hubble diagram of Type Ia Supernovae (SNe Ia) can be naturally explained within this linear evolution. Finally, we calculate the position of the first acoustic peak of the Cosmic Microwave Background (CMB). 

Possibly the strongest motivation to consider the possibility that antimatter presents a negative active gravitational mass comes from the observation of an effective repulsive gravity first observed in 1998 \cite{Riess98,Perlmutter99}, and confirmed by other  independent methods \cite{WMAP1, Astier06}.
Another important motivation for such a symmetric cosmology is provided by the work of Carter \cite{Carter_68} on the Kerr-Newman geometry representing charged spinning black holes. As noted initially by Carter, the Kerr-Newman geometry with the mass, charge and spin of an elementary particle such as an electron bears several of the features expected from the real corresponding particle. In particular, the Kerr-Newman ``electron'' has no horizon, presents automatically the $g=2$ gyromagnetic ratio and has a ring structure with radius equal to half the Compton radius of the electron. 
Additionally, this geometry presents charge and mass/energy reversal symmetries that strongly evoke the CP and T matter-antimatter symmetry \cite{Chardin97, chardin_rax} : when the non-singular interior of the ring is crossed, a second $\mathbb{R}^4$ space is found where charge and mass change sign \cite{Carter_68}. Therefore, starting from an electron with negative charge and positive mass as measured in the first $\mathbb{R}^4$ space, we find in the second space a "positron" with positive charge and negative mass. The relation of the Kerr-Newman geometry to Dirac's equation  and therefore to antimatter has been noted by some authors \cite{Arcos_04, Burinskii08}.

At scales much larger than the matter-antimatter domain scale   -- whose size estimate will be given in next section -- the metric of the Dirac-Milne universe is adequately described by (e.g. \cite{peeblesPPC}):
\begin{equation}
ds^2=c^2dt^2-c^2t^2\left(\frac{dr^2}{1+r^2}+r^2d\Omega^2\right),
\end{equation}
which corresponds to the usual FRLW metric with  the spatial curvature parameter \mbox{$k=-1$} and linear scale factor $a(t)=ct$. The spacetime of this universe is flat, the 4-Riemann tensor of its metric being identically zero. While cosmologists usually consider a flat {\it space} and positively curved spacetime as natural, the Dirac-Milne universe, with its flat {\it spacetime} and open space, appears as another natural geometry.

Temperature evolves in this universe as $T \propto t^{-1}$ instead of $T \propto t^{-1/2}$ in the standard model. This leads to very different timescales at high  temperatures as shown in Table \ref{tab:tab_Tt}.
\begin{table}[htbp] 
\begin{tabular}{r|c|c|}
\cline{2-3}
 &Milne &$\Lambda$-CDM\\
\hline
\multicolumn{1}{|r|}{1 MeV} & 9 $\times 10^7$ s &0.7 s\\
\hline
\multicolumn{1}{|r|}{80 keV} & 1.3 $\times 10^9$ s& 200 s\\
\hline
\multicolumn{1}{|r|}{1 keV} & 1 $\times 10^{11} $s & 1.3 $\times 10^6$ s\\
\hline
\end{tabular}
\caption{\label{tab:tab_Tt}Age of the Universe at a few key temperatures for the Milne and $\Lambda$-CDM models.}
\end{table}
 The present age of the Dirac-Milne universe is simply given by $t_0=1/H_0= 13.9$ billion years with the standard value of $H_0=70\;\rm{km}/\rm{s}/\rm{Mpc}$, nearly identical to the age of the $\Lambda$-CDM model.
This value solves the age problem of the Universe in an elegant way without requiring a Dark Energy component and is therefore another motivation for the Dirac-Milne universe.

The Milne cosmology, interpolation between a strong deceleration of primordial universe and a modern stage of acceleration, has been noted by some authors (\cite{Sethi05} and references therein) to be remarkably concordant. However, their study also exhibits significant problems since deuterium is strongly depleted in the ``simmering'' nucleosynthesis of this cosmology \cite{Sethi99}, while it is difficult to understand why the CMB angular scale could appear at the degree scale.

Primordial nucleosynthesis is one of the most important cosmological tests, providing an observation window back to a temperature of $\approx$ 100 keV. While some tensions exist between the predictions of the Standard Model and the observed abundances of $^{7}$Li  \cite{coc04}, the overall agreement is usually considered impressive, particularly since no satisfying alternative is offered to explain naturally the observed $^{4}$He, $^{7}$Li and D abundances. Some authors \cite{Sethi99, kaplinghat00} have noted, on the other hand, that the Milne universe, despite a primordial nucleosynthesis epoch lasting about 30 years instead of the 3 minutes of the Standard Model, provides adequate abundances for $^{4}$He and $^{7}$Li at the condition that the  baryon to photon ratio $\eta =n_{B}/n_{\gamma}$  has a value $\eta \approx 8 \times10^{-9}$, an order of magnitude larger than the standard value ($\eta \approx 6 \times 10^{-10}$).

\begin{figure}[hbtp]
\includegraphics[width=\columnwidth]{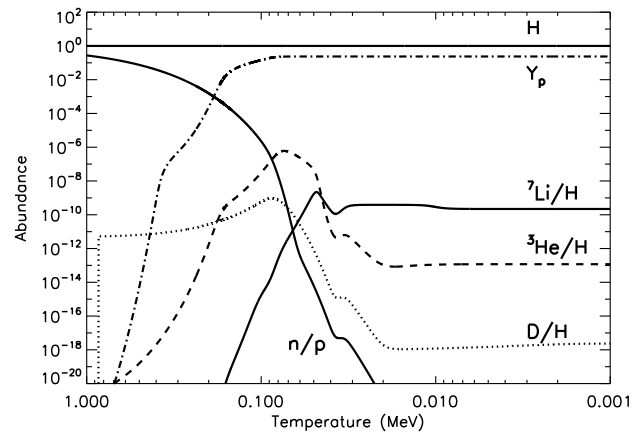}
\caption{\label{fig:mbbn} Abundances of elements produced by homogeneous nucleosynthesis in the symmetric Milne universe for $\eta=8 \times 10^{-9}$.}

\end{figure}

Summarizing this study, the \textquotedblleft simmering\textquotedblright \;  production of $^{4}$He and $^{7}$Li at the observed levels is made possible by the  late  decoupling of weak interactions at a much lower temperature ($\approx$ 80 keV) than the  standard decoupling temperature $\approx$ 1 MeV . Using the latest up-to-date compilations of nuclear reaction rates \cite{nacre,descouvemont2004}, we obtain the abundance curves shown in Fig. \ref{fig:mbbn} for the Dirac-Milne universe. However, as can be seen on this figure, and as noted in \cite{kaplinghat00}, D and $^{3}$He are strongly destroyed and reduced in the Milne universe to unmeasurable levels. Epstein, in a celebrated paper \cite{Epstein76}, demonstrated that deuterium, being very fragile, can only be produced primordially at the observed D/H level of a few $10^{-5}$. The lack of deuterium was therefore considered fatal to the Milne universe \cite{kaplinghat00}.

As noted by Epstein \cite{Epstein76}, the only possibility to produce deuterium by spallation without overproducing other elements and particularly $^{6}$Li, would be to dispose of a nearly monoenergetic beam of energy $\approx$ 400 MeV, which seems quite {\it ad hoc}. But the annihilation of an antinucleon with an $^{4}$He nucleus provides almost exactly this, with typically five pions of average energy 400 MeV. Annihilation at the interface of the matter-antimatter emulsion therefore provides a means to produce deuterium  at observed levels by nucleodisruption, i.e. resulting from the annihilation between an antiproton and a $^4\rm{He}$ nucleus. 
The mechanism of light nuclei production by nucleodisruption in antimatter Big-Bang nucleosynthesis has been extensively  studied in the framework of a standard evolution of the scale factor (see e.g. \cite{Kurki00b, Rehm01}, and references therein).  Residual annihilations at the frontier zone are controlled by diffusion of nucleons. Basically, the only difference between the Dirac-Milne scenario and matter-antimatter scenarios in conventional cosmology comes from the tremendously large timescale available, leading to diffusion lengths orders of magnitude larger.

It is possible to constrain the size of the matter-antimatter emulsion by noting that deuterium is mainly produced by nucleodisruption in $\bar{p}^4$He annihilations. In the Dirac-Milne universe, we show elsewhere \cite{ABL09} that annihilation almost completely stops at a temperature of about 3 eV, i.e. at a redshift $z \approx 10^{4}$ when matter becomes dominant. Before that epoch, essentially all the deuterium produced is subsequently destroyed since the pressure gradient in the expanding universe brings back the light nuclei produced by nucleodisruption in the annihilation zone. Deuterium escaping annihilation is therefore produced during the last Hubble time preceding matter-radiation equality at $z \approx 10^{4}$. In the Dirac-Milne universe, the Hubble time at this redshift is $ \approx 5 \times 10^{13} \;\rm{s}$. At this epoch, the increase of the diffusion coefficient leads to a nearly constant annihilation rate, corresponding to $\approx 7 \;\rm{m}/\rm{s}$ comoving at $z=3\times 10^4$. Therefore, the production of deuterium (and antideuterium) at the observed level of $\approx 3 \times 10^{-5}$ leads to a constraint on the characteristic size of the matter-antimatter emulsion, of the order of $10^{17}\;\rm{m}$ comoving at $z=3\times 10^4$, which may then evolve in structures at the Mpc scale. Note that this inhomogeneous deuterium production leads to an unavoidable dispersion of the observed D/H ratio in QSO measurements of a factor $\sim 2$.

Due to the high baryonic density and to the much larger Hubble time when annihilation stops, it can be checked that radiative processes can thermalize the energy injected in the CMB by the late annihilation down to a redshift of $z\approx 10^4$. At these redshifts, spectral distorsions of the Planckian spectrum are characterized by the dimensionless chemical potential $\mu=1.4 \Delta U/U$, where $\Delta U$ is the energy injected to the radiation energy density $U$. It can be verified that the constraint on $\mu$ from the FIRAS experiment \cite{Fixsen96}: $ \mu < 9\times 10^{-5}$ is met and thus late annihilation should not lead to visible distorsions in the CMB thermal spectrum.

\begin{figure}[hbtp]
\includegraphics[width=\columnwidth]{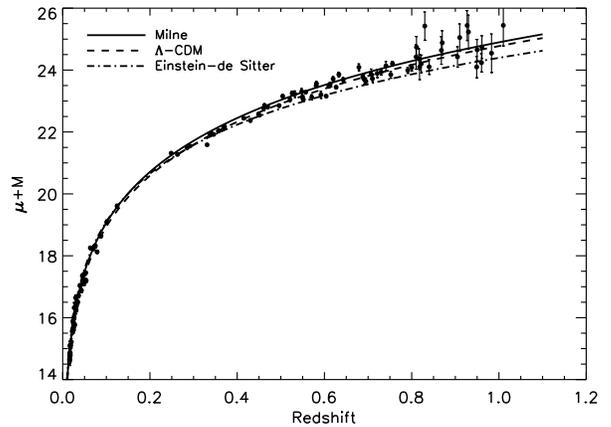}
\caption{\label{fig:hub_diag}Hubble diagram for Dirac-Milne, $\Lambda$-CDM and Einstein-de Sitter models. While the Einstein-de Sitter universe is clearly excluded, the difference between the Milne and  $\Lambda$-CDM universes is much subtler.}
\end{figure}

\begin{figure}[hbtp]
\includegraphics[width=\columnwidth]{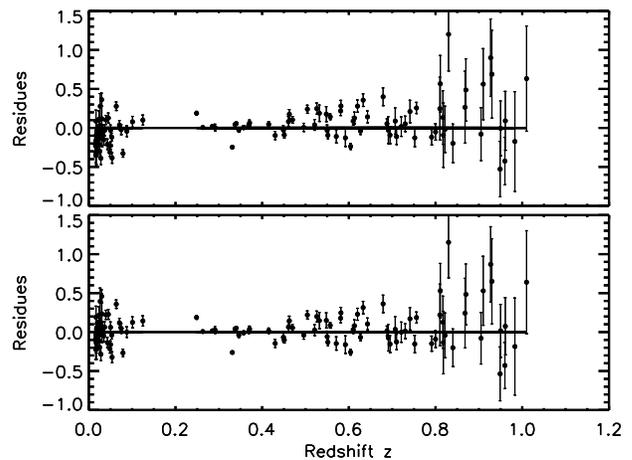}
\caption{\label{fig:residuals}Residuals of the Dirac-Milne (top) and the $\Lambda$-CDM (bottom) adjustment for the first year SNLS data. In this figure, we have plotted the residuals without including the intrinsic dispersion parameter fitted by the SNLS collaboration. It is clear from this figure that the difference between the two models is marginal within the present systematic errors.}
\end{figure}

Another key cosmological test is Type Ia Supernovae (SNe Ia), that provided in 1998 \cite{Riess98, Perlmutter99} a crucial indication that the Hubble diagram of our Universe is incompatible with an Einstein-de Sitter universe. Although these observations are usually interpreted as demonstrating the present acceleration of the expansion, it is important to realize that the Dirac-Milne universe is close to the best fit to the data, without requiring any free parameter.
The Hubble diagram for Dirac-Milne, $\Lambda$CDM and Einstein-de Sitter models, presented in Fig. \ref{fig:hub_diag}, shows that, whereas an Einstein-de Sitter geometry appears clearly excluded, the difference between the $\Lambda$-CDM and the Milne model is much subtler. 
To illustrate further our argument, we have used the data of the first year of the SNLS collaboration \cite{Astier06} and fitted the data to a Milne Universe to get the  corresponding value of the absolute magnitude of SNe Ia which differ from that of $\Lambda$-CDM by $\approx 0.12$ magnitude. The resulting Hubble diagram and the corresponding residues for the $\Lambda$-CDM and the Dirac-Milne model are presented in Fig. \ref{fig:hub_diag} and Fig. \ref{fig:residuals}.
The chi-squared value for SNe Ia fit using all SNLS supernovae with $z > 0.15$ is even {\it lower} for the Dirac-Milne model, without any free parameter. The motivation for this analysis comes from the relatively large uncertainties in the magnitude of the low-z SNe Ia candidates, and the fact that this data sample does not originate from SNLS, but is heterogeneous. 

The Dirac-Milne universe therefore predicts that the absolute luminosity of SNe Ia  is dimmer by $\approx0.12$ magnitude compared to the magnitude fitted in the $\Lambda$-CDM cosmology. Future determinations of this crucial parameter using larger sample of SNe Ia by SNLS and, perhaps even more importantly, improved statistics and reduced systematics on low-$z$ SNe Ia with $0.05 < z < 0.15$ will hopefully allow to distinguish between these two models.

A major result of CMB experiments was to measure precisely the position of the first acoustic peak at the degree scale, which seems to imply that the spatial curvature is nearly zero. In the open spatial geometry of  the Dirac-Milne universe, this position would naively be expected at a much smaller angle since a physical objet at a redshift of $z=1000$ observed under an angle of 1 degree in a $\Lambda$CDM universe would be observed under an angle $\sim$170 times smaller in the Dirac-Milne universe. Let us show that this naive expectation is not met. The angular position of the first peak is defined by the angle under which the sound horizon is seen at recombination: $\theta={r_s(z_*)}/{d_A(z_*)}$, where $r_s(z_*)$ is the sound horizon at the redshift $z_*$ at recombination, and $d_A(z_*)$ is the angular distance  from the present epoch to recombination.

The calculation of the sound horizon requires some care as the mechanisms of sound generation in the Dirac-Milne and $\Lambda$CDM universes differ radically. Contrasting with the generation of sound waves in the Standard Model where inhomogeneities are produced at the epoch of inflation, sound waves in the Dirac-Milne universe are produced by annihilation at the matter-antimatter frontiers. Acoustic waves  propagate in the plasma as long as matter and antimatter are in contact, {\it i.e.}  until $z \approx 10^{4}$ when the baryon loading leads to an almost complete stop of annihilation. After this epoch, the sound horizon simply follows the expansion of the universe. The expression of the angular position of the first acoustic peak then follows:

\begin{equation}
\theta=\frac{180}{\pi}\left(\int^{z_{i}}_{z_{\rm{stop}}}\frac{1}{\sqrt{3}}\frac{dz}{1+z} \right)\times \left(\frac{1}{\sinh \ln(1+z_*)}\right),
\label{peak}
\end{equation}

where $z_{i}\approx 2\times 10^{11}$ is the redshift at which annihilation starts producing coherent sound waves (around $T\sim 40\;\rm{MeV}$), $z_{\rm{stop}}\approx 10^4$ is the redshift at which annihilation almost completely stops, when matter becomes dominant, and $z_*$ is the redshift at decoupling.

Due to the high baryon density in the Dirac-Milne model and the modified evolution of the Hubble rate, the redshift at recombination is found to be  $z^{\rm{Dirac-Milne}}_*\approx 1030$ , whereas the standard value is $z^{\Lambda CDM}_*\approx 1090$.

Calculating the angle of the first acoustic peak using expression (\ref{peak}), we obtain $\theta_{\rm{Dirac-Milne}} \sim 1^\circ $. This remarkable coincidence is quite unexpected and represents a fascinating motivation to study in more detail the Dirac-Milne universe.

In summary, motivated by the observation in our Universe of an effective repulsive gravity, usually ascribed to a Dark Energy component, and by the charge and mass symmetries of the Kerr-Newman solution in General Relativity, we have studied the cosmological properties of the Dirac-Milne universe, a symmetric matter-antimatter universe. We found that this model universe appears to meet surprisingly well the constraints of the main cosmological tests (SNe Ia, nucleosynthesis, age and CMB), without requiring Dark Matter or Dark Energy. Also, this universe does not suffer from the horizon problem, making inflation an unnecessary ingredient.
This first study is sufficiently motivating to calculate in more detail the CMB spectrum resulting from sound waves generated by annihilation at the frontiers of matter-antimatter emulsion. It is important to note that in the Dirac-Milne universe, the Baryon Acoustic Oscillation (BAO) is {\it not} expected to be observed at the scale reported, admittedly at a low statistical significance, by SDSS \cite{Eisenstein05}. This may constitute in the future a crucial test of the Dirac-Milne universe.

In a separate paper \cite{ABL09}, we show how this model universe is able to face the no-go theorem on symmetric matter-antimatter cosmologies studied in \cite{Cohen98}, and study in more details each of these cosmological tests.

 \begin{acknowledgments}
  It is a pleasure to acknowledge fruitful discussions with J. Andrea, E. Armengaud, B. Carter, N. Fourmanoit, J. Fric, K. Jedamzik, E. Keihanen, R. Pain, J. Rich and the members of the SNLS collaboration. Special thanks to A. Coc for allowing us to adapt his nucleosynthesis code. Needless to say, these people are not responsible for the errors present in this, admittedly provocating but hopefully interesting, paper.

\end{acknowledgments}


\end{document}